\documentclass[twocolumn]{revtex4}
\usepackage{epsf}
\usepackage{graphicx}

\begin{document}          

\title{QC-DMRG study of the ionic--neutral curve crossing of LiF}

\author{\"O.~Legeza,\footnote{permanent address: \\
Research Institute for Solid State Physics, H-1525 Budapest, P.\ O.\ Box 49, Hungary}
J. R{\"o}der and B.~A.~Hess}

\affiliation{Chair of Theoretical Chemistry, Friedrich--Alexander University Erlangen--Nuremberg. D-91058 Erlangen, Egerlandstr. 3, Germany}

\date{\today}

\vskip -8pt
\begin{abstract}
We have studied the ionic--neutral curve crossing between the two lowest $^1\Sigma^+$ states of LiF
in order to demonstrate the efficiency of the quantum chemistry version of the density matrix
renormalization group method (QC-DMRG). We show that 
QC-DMRG is capable to calculate the ground and several low-lying excited state
energies within the error margin set up in advance of the calculation,
while with standard quantum chemical methods it is difficult to obtain
a good approximation to Full CI property values at the point of the avoided crossing.
We have calculated
the dipole moment as a function of bond length, which in fact provides a smooth and continuous curve 
even close to the avoided crossing, in contrast to other standard numerical treatments.
\end{abstract}

\pacs{PACS number: 75.10.Jm}

\maketitle

\section{Introduction}

The development of the complete active space self-consistent field (CASSCF) method by Bj\"orn
Roos' group \cite{roos1} was a milestone in the development of electronic-structure
methods effecting a rigorous treatment of electron correlation in molecules. In particular
in the case of nearly degenerate states, i.e., in regions of the potential curve, where 
static correlation prevails, a CASSCF is almost indispensable for a good zero-order
description of the electronic wave function, which usually is subsequently refined by 
methods like MRCI (multi-reference CI) or CASPT2 (second order perturbation theory 
based on a CAS reference state)\cite{anderson1,anderson2}. 
The latter methods focus on dynamic correlation and are mandatory for achieving quantitative 
agreement with experiment. A frequently studied test system for the quantitative calculation
of electron correlation effects in molecules is the ionic--neutral avoided crossing in the 
potential curves of alkaline halogenides, in particular of LiF 
\cite{kahn,botter,werner,spiegelmann1,spiegelmann2,bauschlicher,nakano,finley1,finley2}.
The goal of these benchmark studies is the development of a computational protocol which provides 
a globally accurate description of the system even though the wave functions change dramatically
transversing the avoided crossing. New ab initio methods for the treatment of correlation should 
prove to provide a satisfactory description of the two states in question, and the present
paper is devoted to the application of the quantum-chemistry density matrix renormalization
group (QC-DMRG) to calculate the potential energy curve and the dipole-moment functions in the
vicinity of the ionic--covalent avoided crossing in LiF.

Since its first application on the water molecule \cite{white3,white4} the QC-DMRG algorithm has
become an alternative candidate to treat the electron--electron correlation in atoms and molecules
in a rigorous way. The method has already been applied successfully to calculate the ground state
energy of various molecules containing up to 57 orbitals \cite{mitrushenkov,chan,legeza}, demonstrating
the capability of the method to provide the Full-CI results. The general trend of the accuracy of the
first and second excited states calculated by the method and its strong connection to the structure of
the reduced subsystem density matrix has also been studied in detail on molecules with
various fraction of filling and orbitals \cite{legeza}. Although results of DMRG calculations seem
promising, it is still an open question to determine the stability and performance of the new method for
systems in which the wave function character changes as a function of bond length, especially in the
region of an avoided crossing.                          

In order to provide a benchmark test of the QC-DMRG algorithm for an 
ionic--neutral potential curve crossing, we have calculated the two lowest $^1\Sigma^+$ 
states of LiF.
This system has been subject
of various earlier theoretical studies \cite{kahn,botter,werner}. 
Kahn, Hay and Shavitt \cite{kahn} calculated the four lowest 
$^1\Sigma^+$ states using the configuration-interaction (CI)
methods. The dipole moment function for the lowest $^1\Sigma^+$ states was studied by Werner and Meyer 
\cite{werner} 
using MCSCF wave functions. The remaining discrepancy between their results and the empirical
estimate of Grice and Herschbach \cite{grice} was finally resolved by Bauschlicher and Langhoff who have
employed full configuration interaction calculations to define a protocol using an MRCI approach based on 
carefully constructed CASSCF orbitals,
which was found to satisfactorily represent the FCI potential and dipole moment curves
for the two lowest $^1\Sigma^+$ states of LiF. Problems encountered in the application of CASPT2
\cite{spiegelmann1,spiegelmann2} have been recently resolved on occasion of the development of multi-state CASPT2
\cite{finley1,finley2}.                    

Our aim in this paper is to assess the stability of the QC-DMRG method in the region of an avoided crossing
and show that QC-DMRG provides the FCI energy within an error set up in advance of the
calculation for all bond lengths. 
Taking
the dipole moment as an example, we present a general QC-DMRG procedure to obtain expectation values of
first order-properties for the ground and excited states in molecules using the DMRG method. To the best of our
knowledge, this is the first time that excited-state data and properties have been calculated for a molecule
in the framework of {\em ab initio} QC-DMRG.
 
The setup of the paper is as follows. In Sec.~II we briefly describe the main 
steps of DMRG and  
the details of the numerical procedure used to determine the appropriate ordering and
target state.  
Sec.~III contains the numerical results and analysis of the 
observed trends of the numerical error. The summary of our conclusions 
is presented in Sec.~IV.

\section{Numerical procedure}

Detailed description of the DMRG algorithm can be found in the original papers
\cite{white1,white2,xiang,erick} and its application in the context of quantum 
chemistry is summarized in three recently published papers \cite{mitrushenkov,chan,legeza}. 
Therefore, we present only the most important formulas and definitions that are relevant to 
the question of current interest. 

\subsection{Hamilton operator}

In the context of quantum chemistry a one-dimensional chain that is studied by DMRG 
is built up from the
molecular orbitals that were obtained, e.g., in a suitable mean-field or MCSCF calculation.
The electron--electron correlation
is taken into account by an iterative procedure that minimizes the Rayleigh quotient 
corresponding to the Hamiltonian describing the electronic structure of the molecule,
given by 
\begin{equation}
{\cal H} = \sum_{ij\sigma} T_{ij} c^\dagger_{i\sigma}c_{j\sigma} + 
           \sum_{ijkl\sigma\sigma^\prime} {V_{ijkl} 
           c^\dagger_{i\sigma}c^\dagger_{j\sigma^{\prime}}c_{k\sigma^{\prime}}c_{l\sigma}}
\label{eq:ham}
\end{equation}
and thus determines the full CI wavefunction. 
In Eq.~(\ref{eq:ham}), $T_{ij}$ denotes the matrix elements between orthogonal molecular orbitals \{$\phi_i$\}
of the one-particle Hamiltonian 
comprising kinetic energy and the external electric field of the nuclei,  
while $V_{ijkl}$ stands for the matrix elements of the electron repulsion operator and is defined as

\begin{equation}
V_{ijkl} = \int\! d^3x_1 d^3x_2\, \phi^*_i(\vec x_1) \phi^*_j(\vec x_2) \frac{1}{|\vec x_1-\vec x_2|} \phi_k(\vec x_2)\phi_l(\vec x_1).
\label{eq:vijkl}
\end{equation}

\subsection{Target state}
 
The desired eigenstate of the Hamiltonian that one wants to calculate 
is called the {\em the target state} in the context of DMRG.
The representation of such state in the DMRG procedure
is derived from an arrangement of molecular orbitals in a linear chain, which in turn 
is divided into smaller subsystems and the wave function of the molecule is
built up
from the linear combination of basis states representing these subsystems.            

\begin{figure}
\includegraphics[scale=0.45]{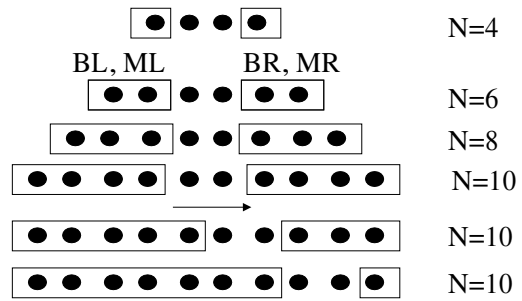}
\caption{The decomposition of the wavefunction to subsystems.}
\label{fig:dmrg}
\end{figure}

According to Fig.~\ref{fig:dmrg} the structure of the subsystems that is called {\em superblock}
configuration is defined as $B_L \bullet \bullet B_R$,
where $B_L$ and $B_R$ represents a block containing $l$ and $r$ molecular orbitals, respectively,
and the $\bullet$ denotes an orbital,
in DMRG terminology a quantum site with $q$ degrees of freedom.
In our case $q$ is equal to four, enumerating the doubly occupied, spin up, spin down and empty 
orbital situation. 
Since the number of molecular orbitals ($N$) is constant, it can be expressed as $l+r+2=N$. 
The $B_L$ block is represented by $M_L$ so-called renormalized block basis states ($M_L \leq q^l$), and each 
block state represents a special distribution of
a given number of electrons with up and down spins on the $l$ orbitals.
The same holds for $B_R$ with $M_R$ block states. 

In order to recover interactions of electrons in orbitals that are far away 
from each other in the chain and 
include their contribution to the correlation energy, 
the size of the left block is increased as long as $l\leq N-3$ and 
the length of the right block is decreased so that $r+l+2=N$ always holds.  
This iterative procedure is repeated in the reverse way as well, until
$r=N-3$ and $l=1$ as is shown in Fig.~\ref{fig:dmrg}. This sweeping procedure 
implies that in each DMRG iteration step the Hamilton operator is constructed on a 
given superblock configuration and the wave function of the molecule is built up
from the linear combination of the subsystem basis states with
coefficients determined by diagonalizing the superblock Hamilton operator.
If $|I\rangle$ and $|J\rangle$ denote basis states for
$B_L \bullet$ and $\bullet B_R$ subsystems, respectively, then the $\beta$th eigenstate is written as                                                                                         
\begin{equation}
\Psi_{\beta} = \sum_{I,J}^{M_L*q,M_R*q} \psi_{I,J}^\beta |IJ \rangle, \qquad |IJ\rangle = |I\rangle \otimes |J\rangle,
\label{eq:psi}
\end{equation}
where $\psi_{I, J}^\beta$ is determined by diagonalization of the superblock
Hamiltonian. In the simplest situation, the target state of the DMRG method that we want to calculate  
is one of the eigenstates of the Hamilton operator and can be usually 
determined with a good accuracy if the eigenstates are well separated. 

As second key ingredient of the DMRG method is the selection of the block states for a subsequent 
iteration step. The $M_L^{\rm new} \leq M_L*q$ block states are chosen from the eigenstates ($\omega_\alpha$) of    
the reduced density matrix of the $B_L \bullet$ subsystem, that is constructed 
for the target state as
\begin{equation}
\rho_{I,I^{\prime}}^\beta = \sum_J \psi_{I,J}^\beta\psi_{I^{\prime},J}^\beta.
\label{eq:rho}
\end{equation}
The error of the truncation procedure is measured by means of the deviation of the
total weight of the selected states from unity which is defined as 
\begin{equation}
TRE=1-\sum_{\alpha=1}^{M_L^{\rm new}} \omega_\alpha.
\label{eq:tre}
\end{equation}
As is was shown in Ref.~\onlinecite{legeza} the accuracy of the DMRG calculation can be
controlled by keeping the truncation error below a given treshold $TRE_{max}$ and 
the number of block states above a minimum value $M_{min}$ which 
avoids the iterative procedure being trapped by local attractors.

It is thus evident, that the construction of the
target state has strong effect on the structure of the subsystem density matrix, on the
block states used during all the subsequent iteration steps and on the 
converging properites of the DMRG method. If the target
state does not contain those basis states that are important to represent the desired
electronic configuration 
at a given bond length, the method can converge to a wrong state by 
loosing the optimum structure of the target state. This has long been confirmed by  
various DMRG studies and the problem can be solved by constructing 
target states by coherent superposition of more eigenstates, 
in which case
\begin{equation}
\Psi_{Target} = \sum_{\beta} \omega_\beta \sum_{I,J}^{M_L*q,M_R*q} \psi_{I,J}^\beta |IJ \rangle.
\label{eq:lin}
\end{equation}
The weights are usually taken as $\omega_\beta=1$ for all $\beta$.

Another possibility to construct block states for the subsequent iteration step is 
first constructing the reduced subsystem density matrices for all eigenstates 
independently and than add them up, in which case the density matrix for an
incoherent target states is defined as 
\begin{equation}
\rho_{I,I^{\prime}} = \sum_\beta \omega_\beta \sum_J \psi_{I,J}^\beta \psi_{I^{\prime},J}^\beta
\label{eq:rho}
\end{equation}
where again, $\omega_\beta=1$ could be introduced. The construction of coherent or incoherent 
superpositions for targeting several excited states simultaneously is the the DMRG
counterpart of the construction of averaged-states orbitals in a CASSCF calculation in order 
to obtain an optimal description of several many-particle states at the same time.

\subsection{Identifying the energy levels}

In the QC-DMRG algorithm block states 
representing a given number of electron with up and down spins are
identified with $N_{up}$ and $N_{down}$ quantum numbers. These are used in order to select out 
the proper combination of block basis states for the total system described in 
a given $m_s = N_{up}^{TOT}-N_{down}^{TOT}$ spin sector. 
In a straightforward application of the DMRG method the first few
eigenvalues of each $m_s$ spin sector are determined and the classification of the singlet, triplet,
quintet etc.\ levels can be accomplished based on the degeneracy of the levels. 
It is, however, often difficult to determine whether two close-lying energy
calculated in two different $m_s$ sectors are two components
of the same level or not. 
This can be resolved by classifying
the block states with respect to the irreducible
representations of as many symmetry groups as possible in order to make use of their direct
product in targeting the right quantum states. 

In our present implementation we use apart from the particle number operator and the $\hat S_z$
operator also the spin reflection operator, which enables us to separate the triplet manifold
from the singlet manifold by targetting the states of appropriate parity of their spin part. 
Obvious extensions in future work would be the implementation of the point-group symmetry of the 
molecule and the $\hat S^2$ operator.

\subsection{Dipole moment function}
 
In the DMRG procedure the expectation value of the one electron operators can be
calculated from the one-particle density matrix according to
\begin{equation}
\langle A \rangle = TR(\rho A),
\label{expectation-values}
\end{equation}
where $A$ is an $N$ by $N$ matrix of the operator for a first-order property, 
in the same one-particle basis that was used to construct
the original $T_{ij}$ and $V_{ijkl}$. 
 
Once the target state was obtained, the one particle subsystem reduced density 
matrix $\rho$ can be formed for any  $B_L \bullet$ configuration as
\begin{equation}
\rho_{ij} = \langle \Psi_{Target} | \sum_\sigma  c_{i\sigma}^\dagger
c_{j\sigma} | \Psi_{Target} \rangle
\end{equation}
where $i$, $j$ denote sites in the left block.  The one-particle density matrix
for the right block is determined in a similar way. If $i$ is in the left block
and $j$ in the right block, then $\rho_{ij}$ is constructed from the 
one-particle operators of the two blocks.

\subsection{Basis states} 

Comparing with the benchmark full-CI calculation of Bauschlicher and Langhoff \cite{bauschlicher} 
we have correlated six electrons of the 
LiF molecule by the QC-DMRG method on a chain built up from 25 CASSCF orbitals. The active space
used to determine the latter contains two $a_1$ orbitals and no $b_1,b_2$ or $a_2$ orbitals. The MOLPRO
program \cite{MOLPRO} was used to prepare    
the $T_{ij}$ and $V_{ijkl}$ matrix elements in this basis. 

The Gaussian basis set used for lithium was derived from the
(9$s$) primitive set of Huzinaga \cite{huzinaga}, supplemented by the (4$p$) functions 
optimized for the $^2P$ state as
in Ref.~\onlinecite{dunning1}. The $s$ and $p$ primitives are contracted (5211) and (31), respectively. 
The F basis set is the
Dunning \cite{Dunning} [$4s2p$] contraction of the Huzinaga \cite{huzinaga} ($9s5p$) primitive set. To
describe $F^-$, a diffuse 2$p$ and a 3$d$ polarization function are added; the exponents are 0.08 and
1.4, respectively.

\subsection{Ordering molecular orbitals}
 
It has become evident that the efficiency and accuracy of the QC-DMRG method depends strongly
on the ordering of the orbitals  
\cite{erick,chan,legeza}. Since this question is still an 
open field of research, we have repeated our previous calculations on H$_2$O, CH$_2$ and
F$_2$ molecules in order to determine a strategy for the best ordering for LiF. 

We have applied various algorithms which reorder the row and column indices
of a sparse matrix in a way such that off-diagonal elements are moved closer to the diagonal or to
one corner of the matrix. Extending the work of Chan and Head-Gordon \cite{chan}, we have included 
a criterion based on 
the Fock operator as well as their originally proposed criterion based on $T_{ij}$. 
Thus we applied the reordering algorithms to the
\begin{equation}
F_{ij} = T_{ij} + \sum_{k\ occupied}{ (4V_{ikkj} - 2V_{ikjk}) }  
\label{eq:fock}
\end{equation}
matrix in order to include some components of the electron--electron interaction term as well. 
In the reordering process, we have neglected terms of $F_{ij}$ below a certain threshold in order
to pick up only the most important components. Of course, this procedure is useful only in the case of 
orbitals differing from Hartree--Fock orbitals, because for the latter $F_{ij}$ is diagonal.

In agreement with the work of Chan and Head-Gordon we have found that a particularly satisfactory 
ordering can be achieved by using the Cuthill-McKee method \cite{cuthill1,cuthill2}.  
As a benchmark test, in case of F$_2$
molecule with $TRE_{max}=10^{-10}$ and $M_{min}=64$ parameter sets a $10^{-6}$ in absolute error     
of the ground state energy was reached with $M_{max}=950$--1100 states, in contrast to our
previous results \cite{legeza} when the same accuracy was reached with 1700--1800 block states. 
In addition, the number of sweeps was reduced by a factor of two.  Similar results were
found in the other test cases as well. Thus we can conclude that moving the off-diagonal elements
closer to the diagonal with the Cuthill-McKee algorithm 
reduces the number of block states and accelerates the speed of convergence. 
It is worth to note, that this protocol is only expected to give good results if a mean-field
description of the system is reasonable, i.e., in weakly correlated cases.

Based on our observations, we have applied the Cuthill-McKee method on the one-electron 
$F_{ij}$ interaction matrix to determine the ordering configuration for LiF for  
all the different bond lengths. 
We have found the same ordering at both sides of the avoided crossing; the orbital ordering
which we used in our calculations is detailed in Table~\ref{ordering}.

\subsection{Controlling numerical accuracy}

It has been shown in Ref.~\onlinecite{legeza} that the accuracy of the QC-DMRG method can 
be controlled by method which we called
dynamic block state selection (DBSS) approach, whereby the 
desired accuracy can be predetermined in advance of the calculations. Based on the Full-CI energies
calculated at several points on the potential curve we can infer that the  
smallest value of the energy gap close to the avoided crossing  
is in the order of $10^{-3}$, thus it is 
sufficient to obtain an accuracy of $10^{-4}$ in the absolute error of the energy levels. In case 
of the LiF molecule represented with the given basis states this means a relative error of 
$10^{-6}$, thus the parameter that controls the maximum value of the truncation error 
has been set to $10^{-7}$. For the minimum number of block states ($M_{min}$) a few 
hundred states would be sufficient. 

\begin{figure}
\includegraphics[scale=0.45]{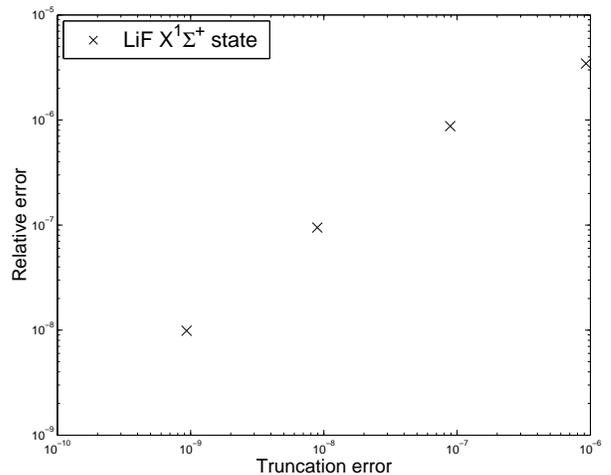}
\caption{Relative error of the ground state energy as a function of the truncation error.}
\label{fig:lif_scale}
\end{figure}

We have made several test calculations with the above parameter set and found that the relative error
indeed converged to $10^{-6}$. As an example of the linear relationship between the relative error
and the truncation error we have plotted our results obtained for various values of $TRE_{max}$ parameter
close to the avoided crossing ($R=11.5$ a.u.) in Fig.~\ref{fig:lif_scale}. 

On the other hand, in order to provide a better benchmark test, we have also used a smaller 
value of $TRE_{max}$ for most of our calculations.  
We have set $TRE_{max}$ to $10^{-9}$ which ensures 
an absolute error of $10^{-7}$ for the lowest energy state of a given $m_s$ spin sector. 
The accuracy of the excited state energies, on the other hand, can depend strongly on other 
parameters and the construction of the target state which will be examined in detail 
in the next section.

\section{Numerical results}

We have carried out a detailed DMRG study of the absolute error of the
ground and several excited state energies and the dipole moment  
as a function of bond length, focusing on the avoided crossing, 
where the wavefunction and also the convergence properties of DMRG is expected to change rapidly.

\subsection{Ground state}

According to our previous study \cite{legeza} first we ran a calculation with 
$M_{min}=164$ block states and repeated the same calculation with $M_{min}=64$. 
We have found that using the Cuthill-McKee reordered orbital configuration 
the desired accuracy was achieved with the smaller $M_{min}$ value as well, thus we
have carried out all of our calculations with the latter value. This in fact sped up the
calculation significantly reducing the computational time to an order of an hour on
a standard PC running at 1.5 GHz.

\begin{figure}
\includegraphics[scale=0.45]{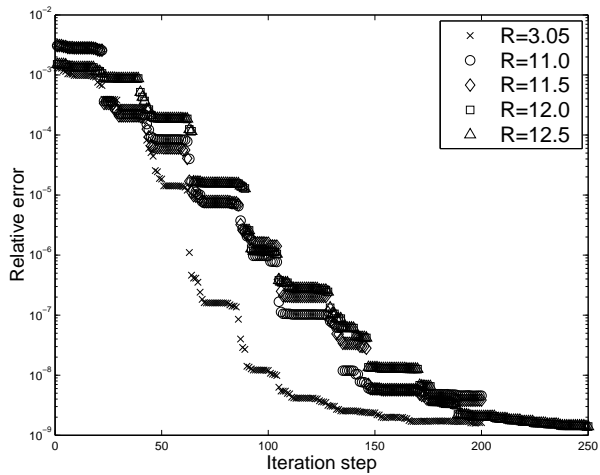}
\caption{The relative error of the ground state energies a function of iteration steps 
at various bond lengths}
\label{fig:energy}
\end{figure}

We have found that the desired accuracy of the ground state was achieved for all 
bond lengths, even close to the avoided crossing. As an example we have plotted
in Fig.~\ref{fig:energy}
the relative error of the ground state energy as a function of the DMRG 
iteration steps at various bond lengths.
It can be seen that the relative error of the ground state energy converges to $10^{-9}$ in all cases.  
The speed of convergence, however, is slower closer to the avoided crossing  
as expected be because of the smaller gap and longer coherence length.
We have also found that the largest numbers of the block states ($M_{max}$) 
generated dynamically fell in the range of 150--200 at the equilibrium point, while it
was slightly larger (400--500) close to the avoided crossing.

\subsection{Excited states}

In order to determine higher-lying excited states, we have used the spin flip
operator to separate the singlet and triplet components. We have found that the 
first excited state ($^3\Pi$) is the lowest state of the even spin symmetry sector at
small bond distances and exhibits the 
same convergence property as the ground state. Targeting the second level of the 
odd spin symmetry sector provided the $^1\Pi$ level at the equilibrium point at 
$R=3.05$ up to six digits, as expected. We have also targeted higher-lying 
levels to obtain the second $^1\Sigma^+$ state, but the convergence of the 
method was much slower.

\begin{figure}
\includegraphics[scale=0.45]{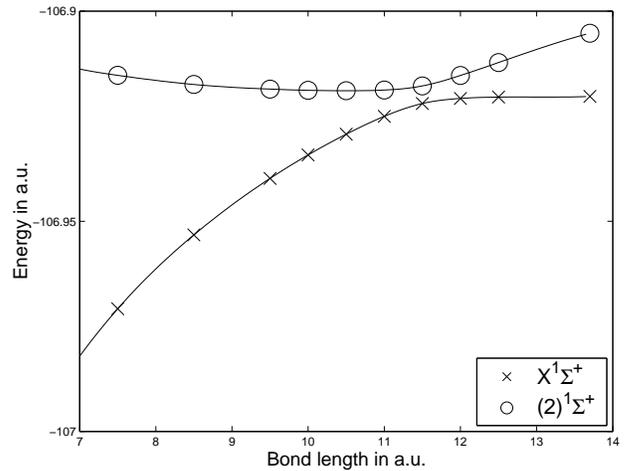}
\caption{The calculated potential curve in the vicinity of the avoided crossing. The DMRG results agree
up to six digits with the Full CI results.}
\label{fig:lif_pot}
\end{figure}

Targeting the second level of the even spin sector for increasing bond lengths, we have found
that the target state can be lost when $R$ is close to the avoided crossing. 
This is, in fact, not unexpected, and we made use of coherent and incoherent target states 
in order to resolve the problem. 
In addition, to provide better stability of DMRG it is advantageous to use larger values 
of $M_{min}$, thus we have set it to 200--300. 
Using the incoherent target states method, the $^1\Sigma^+$ state was also determined within 
an accuracy of $10^{-7}$ for all bond lengths. 
Our result for  
the obtained potential curve is plotted on Fig.~\ref{fig:lif_pot}, which agrees up to
six digits with the Full-CI result, also given in Table~\ref{table:energy}. 
Bauschlicher and Langhoff have pointed out that a 
careful selection of the CASSCF space used
in the determination of the orbitals for the MRCI treatment is mandatory in order to get
close agreement with FCI results. By contrast, we have found that routine methods 
for the determination of the orbital basis are adequate in our DMRG treatment. 
Indeed, Hartree--Fock orbitals yielded practically the same results, corroborating the view that 
DMRG treatments can be effectively converged to the Full CI result.
We have also calculated the energy levels of the $m_s=\pm1,2$ sectors and found that the 
the residual splitting of the $m_s=0,1,-1$ components of the
triplet level was as low as $10^{-10}$ a.u.

\subsection{Dipole moment function}

\begin{figure}
\includegraphics[scale=0.45]{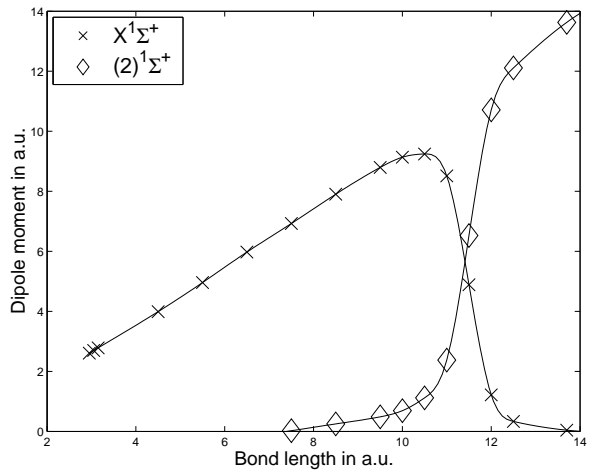}
\caption{The ground and excited states dipole moment as a functions of the bond length.}
\label{fig:dipole1}
\end{figure}

We have calculated the one-particle density matrix and the dipole moment for the
various bond lengths. The dipole moment function is obtained 
as a smooth curve close to the Full-CI result. Note that especially the determination of the
dipole moment curve in the vicinity of the avoided crossing proved very difficult in case
of approximate methods of standard quantum chemistry. No special parameters are used in the 
DMRG calculation, i.e., it may be used as a `black box' method, whereas considerable skill is required
to choose the appropriate CASSCF parameters necessary to obtain satisfactory orbitals for an MRCI treatment.
The absolute error of the dipole moment function was $10^{-5}$ at the equilibrium point and the
largest error of the order of $10^{-3}$ was found at $R=11.5$ a.u.
Our results for the ground and excited states are shown in Fig.~\ref{fig:dipole1} and
detailed in Table~\ref{table:dipole}.

\subsection{The stability of structure of the wavefunction}

\begin{figure}
\includegraphics[scale=0.45]{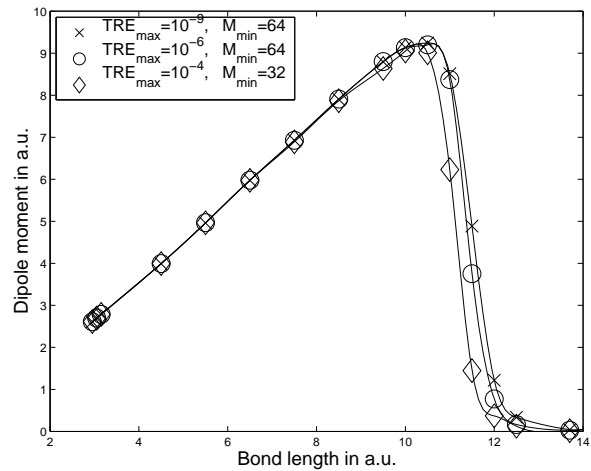}
\caption{The ground state dipole moment as a function of the bond length calculated 
for the different values of $TRE_{max}$ parameter.}
\label{fig:dipole2}
\end{figure}

Within the framework of the configuration interaction method 
the number of Slater
determinants included in the treatment is increased systematically 
in order to achieve a better accuracy. Thus the wavefunction is improved by adding in turn 
single, double and higher excitations. On the other hand,
the coupled cluster procedure includes higher excitations in a different fashion, adding 
excitations of a
very limited kind (namely the unlinked clusters) at a very early stage of approximation. 
By contrast,
DMRG takes into account excitations even of higher excitation classes
at the very beginning, picking up the 
most important ones by minimizing the energy. This means that if 
the accuracy of the calculation is lowered, the structure of the wave function is retained in essence.
To demonstrate this, we have recalculated the 
ground state energies and the dipole moment with $TRE_{max}=10^{-6}$ and found
that the dipole moment function is still a continuous function even close
to the avoided crossing. The relative error
of the ground state energy was found to converge to $10^{-5}$ and the dipole 
moment to $10^{-4}$ at $R$=11.5 a.u. Repeating the same calculation by cutting the 
parameters drastically and using $M_{min}=32$ and 
$TRE_{max}=10^{-4}$, the dipole moment is still a continuous curve but it 
deviates more significantly from the FCI results especially
close to the avoided crossing (Fig.~\ref{fig:dipole2}).
This simply means that some of the important components of the 
wave function are lost at the very beginning,
which is of particular importance at the avoided crossing. 

\section{Summary}

We have studied the low-lying energy spectrum of LiF molecule by the 
quantum chemistry version of DMRG in order to demonstrate the efficiency 
of the method in the vicinity of an avoided crossing.  
Besides calculating the ground and several excited states
energies, we have obtained the dipole moment function as well. Several
technical aspects of the QC-DMRG method were further studied.     
We have extended the application of
the dynamic block state approach and the study of optimal ordering configuration of 
molecular orbitals.  

We have found that reordering the molecular orbitals with the 
Cuthill-McKee algorithm applied on the $T_{ij}$ matrix elements or the Fock matrix
the same accuracy compared to an ordering based on the orbital energy
can be achieved with a smaller minimum number of block states ($M_{min}$).
The QC-DMRG procedure selects out a significantly smaller number of block states, $M_{max}$, 
and the number of iteration steps is reduced. In addition, we
have shown the efficiency and stability
of the dynamic block state selection approach which in fact reduces
the computational time significantly.

It turns out that the desired accuracy of the lowest states 
of each spin symmetry sector
can be achieved for all bond lengths by targeting that particular state.
On the other hand, calculating higher-lying levels close to the avoided crossing,
coherent or incoherent superpositions of target states have to be constructed 
in order to provide stability and convergence. Since this has no significant 
effect on calculations far away from the avoided crossing where the energy gap is large,
this method can be used for the whole range of bond lengths.

The dipole moment calculated by DMRG can 
also be determined with good accuracy for all bond lengths and in fact
it is a continuous curve even close to the avoided crossing.  
We have found that adjusting the desired accuracy of the calculation 
affects the accuracy the dipole moment, but the dipole moment
function remains a continuous curve. This demonstrates the power of DMRG in
quantum chemistry, since the structure of wavefunction does not change significantly
for lower accuracy, namely the most important configurations are picked up even
with lower accuracy.

As far as the technical questions are considered, using reordered orbitals and the
DBSS approach QC-DMRG calculation time was found to be within the
order of magnitude of the FCI calculation if point-group symmetries
are not used. Although the point group symmetry is not built in our present
implementation of the DMRG method, 
the structure of the $V_{ijkl}$ matrix reflects the symmetry, and
the number of nonzero matrix elements is reduced significantly. 

\begin{acknowledgements}
This research was supported in part by the Fonds der Chemischen Industrie and   
the Hungarian Research Fund (OTKA) Grant No.\ 30173 and 32231. The authors also
thank Christian Kollmar for providing FCI results for the dipole moment 
function.   
\end{acknowledgements}

\begin{widetext}
\begin{center}
\begin{table}
\begin{tabular}{|l|l|l|l|l|l|l|l|l|l|l|l|l|l|l|l|l|l|l|l|l|l|l|l|l|}
\hline
12 & 11& 10& 9& 8& 7& 6& 5& 4& 3& 1& 2& 18& 17& 16& 15& 13 & 14& 24& 23& 22& 21& 19& 20& 25
\\
\hline
\end{tabular}
\vskip .1 truein 
\caption{Orbital ordering employed in this work}
\label{ordering}
\end{table}                    

\begin{table}
\begin{tabular}{ccccc}
         &                  &                    &                     &         \\
  R      &    DMRG          & DMRG               &    FCI              & FCI        \\
 a.u.    &    $E_{GS}$      & $E_{(2)^1\Sigma^+}$          &      $E_{GS}$       &  $E_{(2)^1\Sigma^+}$\\\hline 
 2.95    &    $-107.111724$   &     $-106.797843$    &      $-107.111726$    &      $-106.797846$  \\
 3.05    &    $-107.113052$   &     $-106.806390$    &      $-107.113055$    &      $-106.806392$  \\
 5.50    &    $-107.024654$   &     $-106.873564$    &      $-107.024656$    &      $-106.873565$  \\ 
 6.50    &    $-106.994852$   &     $-106.892601$    &      $-106.994853$    &      $-106.892602$  \\ 
 7.50    &    $-106.970816$   &     $-106.915277$    &      $-106.970817$    &      $-106.915277$  \\ 
 8.50    &    $-106.953232$   &     $-106.917457$    &      $-106.953232$    &      $-106.917457$  \\ 
 9.50    &    $-106.939803$   &     $-106.918556$    &      $-106.939803$    &      $-106.918556$  \\ 
10.00    &    $-106.934216$   &     $-106.918837$    &      $-106.934216$    &      $-106.918838$  \\ 
10.50    &    $-106.929276$   &     $-106.918948$    &      $-106.929277$    &      $-106.918948$  \\ 
11.00    &    $-106.925037$   &     $-106.918782$    &      $-106.925037$    &      $-106.918782$  \\ 
11.50    &    $-106.921985$   &     $-106.917803$    &      $-106.921985$    &      $-106.917804$  \\ 
12.00    &    $-106.920778$   &     $-106.915314$    &      $-106.920779$    &      $-106.915314$  \\ 
12.50    &    $-106.920455$   &     $-106.912236$    &      $-106.920455$    &      $-106.912239$  \\ 
13.70    &    $-106.920295$   &     $-106.905273$    &      $-106.920295$    &      $-106.905274$  \\   
\end{tabular}
\vskip .1 truein 
\caption{The ground and first excited state energies in atomic units as functions of bond length.} 
\label{table:energy}
\end{table}                    

\begin{table}
\begin{tabular}{ccccc}
         &                  &                    &                     &         \\
  R      &    DMRG          & DMRG               &    FCI              & FCI        \\
 a.u.    &    $\mu_{GS}$      & $\mu_{(2)^1\Sigma^+}$          &      $\mu_{GS}$       &  $\mu_{(2)^1\Sigma^+}$\\\hline 
 2.95    & $2.609571$    &$-0.461813$   &    $2.609585$   &    $-0.461810$ \\
 3.05    & $2.697383$    &$-0.462597$   &    $2.697402$   &    $-0.462597$ \\ 
 5.50    & $4.959323$    &$-0.528943$   &    $4.959400$   &    $-0.528936$ \\ 
 6.50    & $5.975189$    &$-0.161147$   &    $5.975169$   &    $-0.161177$ \\ 
 7.50    & $6.926324$    &$ 0.024134$   &    $6.926329$   &    $ 0.024112$ \\ 
 8.50    & $7.903448$    &$ 0.250507$   &    $7.903463$   &    $ 0.250889$ \\ 
 9.50    & $8.795699$    &$ 0.485936$   &    $8.795784$   &    $ 0.485905$ \\ 
10.00    & $9.136214$    &$ 0.689593$   &    $9.136292$   &    $ 0.689562$ \\ 
10.50    & $9.243327$    &$ 1.117497$   &    $9.243389$   &    $ 1.117521$ \\ 
11.00    & $8.513409$    &$ 2.375437$   &    $8.513886$   &    $ 2.376136$ \\ 
11.50    & $4.886492$    &$ 6.524977$   &    $4.889316$   &    $ 6.523027$ \\ 
12.00    & $1.218095$    &$10.711791$   &    $1.218455$   &    $10.711489$ \\ 
12.50    & $0.333931$    &$12.110902$   &    $0.334032$   &    $12.110658$ \\ 
13.70    & $0.042024$    &$13.628204$   &    $0.042074$   &    $13.628195$ \\              
\end{tabular}
\vskip .1 truein 
\caption{The ground and first excited state dipole moments in atomic units as functions of bond length.} 
\label{table:dipole}
\end{table}                    
\end{center}
\end{widetext}


\end{document}